\documentclass[aps,prb,twocolumn,showpacs,amsmath,amssymb,unsortedaddress]{revtex4-1}
\usepackage{graphicx}
\usepackage{longtable}
\usepackage{ulem}
\usepackage{color}

\begin{document}
\title{Influence of post-deposition annealing on the chemical states of crystalline tantalum pentoxide films}

\author{Israel Perez} 
\email[Contact Author: ]{cooguion@yahoo.com}
\affiliation{National Council of Science and Technology (CONACYT)-Institute of Engineering and Technology, Universidad Aut\'onoma de Ciudad Ju\'arez, Av. del Charro 450 Col. Romero Partido, C.P. 32310, Ju\'arez, Chihuahua, M\'exico}
\author{V\'ictor Sosa} 
\affiliation{Applied Physics Department, CINVESTAV Unidad M\'erida, km 6 Ant. Carretera a Progreso, A.P. 73, C.P. 97310 M\'erida, Yucat\'an, M\'exico}
\author{Fidel Gamboa}
\affiliation{{\it Applied Physics Department, CINVESTAV Unidad M\'erida, km 6 Ant. Carretera a Progreso, A.P. 73, C.P. 97310 M\'erida, Yucat\'an, M\'exico}}
\author{Jos\'e Trinidad Elizalde Galindo} 
\affiliation{Institute of Engineering and Technology, Universidad Aut\'onoma de Ciudad Ju\'arez, Av. del Charro 450 Col. Romero Partido, C.P. 32310, Ju\'arez, Chihuahua, M\'exico}
\author{Jos\'e L. Enr\'iquez Carrejo} 
\affiliation{Institute of Engineering and Technology, Universidad Aut\'onoma de Ciudad Ju\'arez, Av. del Charro 450 Col. Romero Partido, C.P. 32310, Ju\'arez, Chihuahua, M\'exico}
\author{Rurik Farias} 
\affiliation{Institute of Engineering and Technology, Universidad Aut\'onoma de Ciudad Ju\'arez, Av. del Charro 450 Col. Romero Partido, C.P. 32310, Ju\'arez, Chihuahua, M\'exico}
\author{Pierre Giovanni Mani Gonz\'alez} 
\affiliation{Institute of Engineering and Technology, Universidad Aut\'onoma de Ciudad Ju\'arez, Av. del Charro 450 Col. Romero Partido, C.P. 32310, Ju\'arez, Chihuahua, M\'exico}
\date{\today}

\begin{abstract}
We investigate the effect of post-deposition annealing (for temperatures from 848 K to 1273 K) on the chemical properties of crystalline Ta$_2$O$_5$ films grown on Si(100) substrates by radio frequency magnetron sputtering. The atomic arrangement, as determined by X-ray diffraction, is predominately hexagonal ($\delta$-Ta$_2$O$_5$) for the films exposed to heat treatments at 948 K and 1048 K; orthorhombic ($\beta$-Ta$_2$O$_5$) for samples annealed at 1148 K and 1273 K; and amorphous for samples annealed at temperatures below 948 K. X-ray photoelectron spectroscopy for Ta $4f$ and O $1s$ core-levels were performed to evaluate the chemical properties of all films as a function of annealing temperature. Upon analysis, it is observed the Ta $4f$ spectrum characteristic of Ta in Ta$^{5+}$ and the formation of Ta-oxide phases with oxidation states Ta$^{1+}$, Ta$^{2+}$, Ta$^{3+}$, and Ta$^{4+}$. The study reveals that the increase in annealing temperature increases the percentage of the state Ta$^{5+}$ and the reduction of the others indicating that higher temperatures are more desirable to produce Ta$_2$O$_5$, however, there seems to be an optimal annealing temperature that maximizes the O\% to Ta\% ratio. We found that at 1273 K the ratio slightly reduces suggesting oxygen depletion.
\end{abstract}

\maketitle
\section{Introduction}
In the last years tantalum pentoxide (Ta$_2$O$_5$) has been considered as a good candidate for technological applications such as gas detector, insulator, catalyst, and storage capacitor \cite{tkaga91a, kwkwon96, cchaneliere98a}. This is due to its high dielectric constant, high refractive index ($n=2.18$ at $\lambda=550 \;\text{nm}$), and a wide optical band gap of $\sim$4.0 eV ~\cite{cchaneliere98b,rhdennard74a,sshibata96a,eatanassova99a}. Ta$_2$O$_5$ crystallizes in several polymorphs depending on the synthesis methods and the temperatures in post-deposition heat treatments. For annealing temperatures below 873 K, Ta$_2$O$_5$ is amorphous but for temperatures between 873 K and 1523 K two crystalline phases show up, namely: hexagonal ($\delta$ phase) and orthorhombic ($\beta$ phase) \cite{jdkruschwitz97a,cchaneliere99a,spgarg96a,ktjacob09a}. 

The chemical, physical, and structural properties of Ta$_2$O$_5$ films deeply depend on the fabrication processes. In the last years, a great number of methods such as PLD, rf sputtering, ion assisted deposition, and EBE have been used to grow Ta$_2$O$_5$ films \cite{eatanassova95a,hshinriki91a,skamiyama93a,gqlo93a,ykuo92a,ndonkov11a}. In general, the oxygen to tantalum ratio is not stoichiometric and the chemical composition depends strongly on the methods employed. To ensure full oxygenation of the films, post-deposition annealing is carried out \cite{tdimitrova01a,sjjwu09a,dcristea13a}. There has been evidence of crystalline phase transitions from hexagonal to orthorhombic as the annealing temperature increases \cite{iperez17a}. As a result of this, significant changes on the physical and chemical properties of the samples are expected. Despite that there is wide number of works studying the chemical properties of amorphous Ta$_2$O$_5$ films \cite{ndonkov11a,eatanassova98a,ttsuchiya11a,scwang11a,svjchandra10a}, to the best of our knowledge, there a few works focused on studying these properties for the crystalline phases. 

In this work we investigate the annealing temperature dependence of the chemical properties in crystalline Ta$_2$O$_5$ films deposited on Si substrates by RF magnetron sputtering. For this purpose we prepared several Ta films and exposed them to annealing temperatures ranging from 848 K to 1273 K. Their crystalline structure was evaluated with XRD (X-ray diffraction) and the chemical properties such as atomic concentration, chemical states, and atomic bonding were studied by X-ray photoelectron spectroscopy (XPS). 
 
\section{Experimental}
\subsection{Film growth and annealing}
Six amorphous Ta films were deposited at room temperature on Si(100) substrates by the RF magnetron sputtering technique. The deposition took place in a vacuum chamber with a base pressure of $6.6 \times 10^{-5}$ mbar. Argon gas (99.9\% purity) was flushed into the chamber to obtain a working pressure of ($2.6\pm 0.1)\times10^{-2}$ mbar. Before deposition the substrates were cleaned up by several baths of distilled water, acetone, and ethanol. In order to eliminate the native oxide layer on the target a 5 min-presputtering was conducted before deposition. The deposition was carried out using a 2.5 inch-Ta target with 99.95\% purity and a sputtering power of 120 W; resulting in a deposition rate of 2.8 \AA$\cdot$s$^{-1}$ and a thickness for all films of 2.4 $\mu$m. The target-to-substrate distance was about 12 cm and during film growth no oxygen was flushed into the chamber. To induce the desired crystalline phase, i.e., Ta$_2$O$_5$, five films were exposed to post-deposition heat treatments in air for 1 h at different annealing temperatures, namely, (848, 948, 1048, 1148, 1273) K using a Thermo Scientific Thermolyne cylindrical furnace (model F21135). To associate the annealing temperature to the samples, the films were labeled F848, F948, F1048, F1148, and F1273. The remaining film, labeled F298, was kept for future reference and was not subjected to any heat treatment. Table \ref{details} gives the annealing temperature $T_{ann}$ and the crystalline phase of the films.
\begin{table}[t!]
	\centering 
	\caption{Annealing temperature $T_{ann}$ and determined crystalline phase for the films.}
	\label{details}
	\begin{tabular}{ccc}
		\hline
		Film   & $\frac{T_{ann}}{K}$  & Phase  \\ \hline
		F298  & 298 &  Amorphous \\
		F848  & 848&  Amorphous \\
		F948 & 948   &$\delta$ \\
		F1048 & 1048  & $\delta$  \\
		F1148 & 1148  & $\beta$  \\
		F1273  & 1273 & $\beta$  \\
		\hline
	\end{tabular}
\end{table}

\subsection{Characterization}
A Siemens diffractometer model D-5000 with Cu $K_{\alpha_1}$ radiation ($\lambda=1.5406$ \AA) was used to evaluate the atomic structure of the samples. Steps of 0.02$^\circ$ with a time per step of 3 s and operating parameters of 34 kV and 25 $\mu$A were used to obtain the XRD patterns. A Thermo Scientific K-Alpha XPS spectrometer with an Al K$_{\alpha}$ X-ray source set to 12 kV and 40 W was used to analyze the chemical properties of the films. For the XPS surveys and scans we used steps of 1 eV and 0.1 eV, respectively; the beam spot had a diameter of 400 $\mu$m and made an angle relative to the sample of 30$^\circ$. Chemical properties were assessed measuring the Ta $4f$ and O $1s$ core-levels. The atomic composition, chemical states, and atomic bonding were determined by deconvolution of the Ta $4f$ and O $1s$ spectra using Shirley-Sherwood background and Voigt functions [Gaussian $\sigma=(1.43, 0.9)$ eV and Lorentzian $\gamma=0.02$ eV] as implemented in the AAnalizer software \cite{aherrerag14a}. 

\section{Results and discussion}
\subsection{Crystalline Structure}
\begin{figure}[t!]
	\begin{center}
		\includegraphics[width=8.4cm]{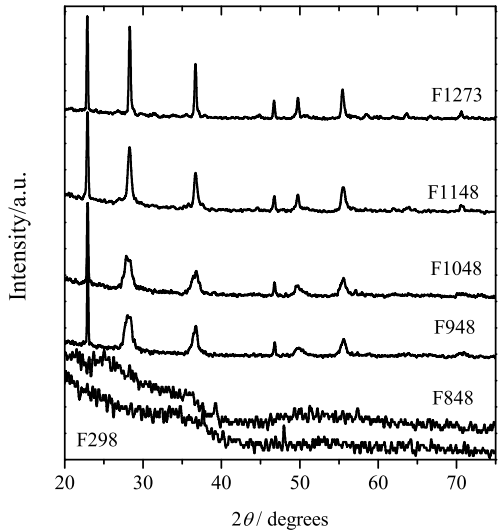}
		\caption{X-ray diffraction patterns for the crystalline films. The patterns for the samples F948 and F1048 are indexed to the $\delta$ phase and the patterns for F1148 and F1273 are indexed to the $\beta$ phase of Ta$_2$O$_5$. Samples F298 and F848 are amorphous.}
		\label{drxS16S61000C}
	\end{center}
\end{figure}

To evaluate the crystalline structure of our samples we performed XRD measurements. Samples F298 and F848 showed no diffraction patterns, indicating a disordered atomic structure (see Fig. \ref{drxS16S61000C}). The rest of the samples exhibit a crystalline structure. 
For the sake of phase indexation we compared our measurements with several references from the Powder Diffraction File (PDF) dababase under the International Centre for Diffraction Data (ICDD) implemented in the software of the diffractometer. From a list of at least six PDFs (even among these references we spotted slight discrepancies) the films F948 and F1048 can be indexed to the hexagonal phase $\delta-$Ta$_2$O$_5$ with PDF 00-019-1299; lattice parameters $a=b=3.6240$ \AA, $c=3.8800$ \AA, $\alpha=\beta=90^\circ$, and $\gamma=120^\circ$; spatial group $P6/mmm$ \cite{afukumoto97a,ynwu11a}; whereas the films F1148 and F1273 can be indexed to the orthorhombic phase $\beta-$Ta$_2$O$_5$ with either PDF 00-025-0922; lattice parameters $a=6.1980$ \AA, $b=40.2900$ \AA, $c=3.8880$ \AA, and $\alpha=\beta=\gamma=90^\circ$; spatial group $P2_12_12$; or PDF 01-089-2843; lattice parameters $a=6.2000$ \AA, $b=3.6600$ \AA, $c=3.8900$ \AA, and $\alpha=\beta=\gamma=90^\circ$; spatial group $Amm2$. 
\begin{figure}[t!]
	\begin{center}
		\includegraphics[width=8.4cm]{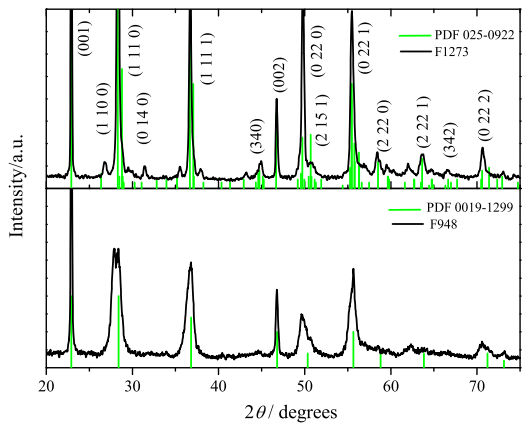}
		\caption{Phase indexation of our samples. Samples F1148 and F1273 can be indexed to PDF 025-0922 whereas films F948 and F1048 can be indexed to PDF 0019-1299.}
		\label{fig2}
	\end{center}
\end{figure}
In Table \ref{details} the phases for all films are summarized. Figure \ref{fig2} shows the best matches; there we can see that the pattern of F1273 shows a much richer pattern than the pattern due to F948. It is evident that F1273 exhibits a series of small reflections along the whole pattern that match the reference pattern PDF 025-0922. It is important to underline that so far there has not been a consensus regarding the spatial symmetry of these two phases, and, due to the overlapping of several peaks, one cannot rule out the coexistence of both phases \cite{iperez17a,sperezw16a,jykim14a,shlee13a,jlee14a,zhelali14a,yguo15a,jykim15a,yyang18a}. To the best of our knowledge, at least 12 spatial groups has been proposed and we do  think it is not worth computing the lattice parameters or conducting a Rietveld refinement, since these depend on the particular choice of the group symmetry. The precise determination of both the crystalline structure and the spatial symmetry are beyond the scope of the present investigation.

The crystallite size $D$ was estimated using the so-called Scherrer relation
\begin{equation}
\label{sch}
D=\frac{K\lambda}{\Gamma \cos\theta},
\end{equation}
where we used $K=0.9$, $\lambda=1.5406\;\text{\AA}$  and $\Gamma$ is the full width at half maximum. For our calculations we used the peak at $2\theta=28.3^\circ$ and found that $\Gamma=$(0.79, 0.69, 0.52, 0.27)$^\circ$. With these values we found the size to be $D=$(10, 12, 16, 30) nm, respectively.

It is worth mentioning that Ta$_2$O$_5$ is the most stable oxide of Ta; other oxides such as TaO, Ta$_2$O, TaO$_2$, Ta$_2$O$_3$ are difficult to synthesize as pure phases (except for TaO$_{x}$) and usually appear as contamination \cite{spgarg96a}. Furthermore, most suboxides are crystalline, except for Ta$_2$O$_3$ and TaO$_x$ that are amorphous. This issue will become important for our forthcoming discussion.

\subsection{Chemical properties}
\begin{figure}[t!]
	\begin{center}
		\includegraphics[width=8.4cm]{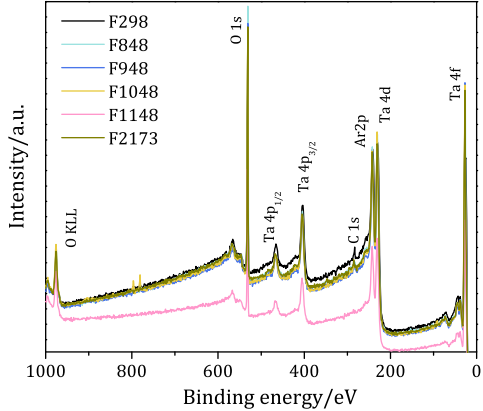} 
		\caption{XPS surveys for all samples. Some degree of carbon contamination in all samples is observed.}
		\label{xpssurvey}
	\end{center}
\end{figure}

The chemical properties of our samples were studied by XPS analysis. The binding energy of all spectra for Ta $4f$ and O $1s$ core-levels was calibrated at 532 eV using as reference the oxygen peak. During the analysis, we also spotted traces of carbon contamination in all films as revealed by the surveys in Figure \ref{xpssurvey}. 

\begin{figure}[t!]
	\begin{center}
		\includegraphics[width=8.4cm]{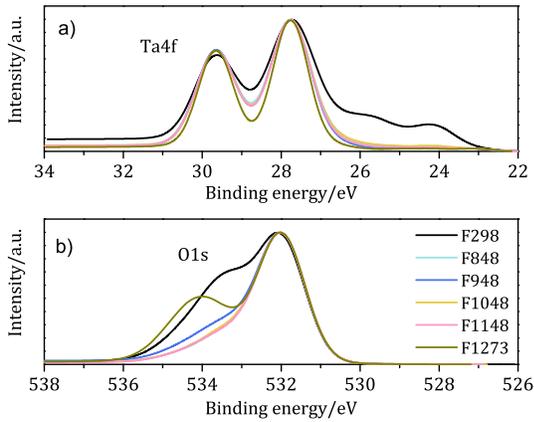} 
		\caption{Ta $4f$ (a) and O $1s$ (b) core-levels for the six films.}
		\label{xpstao}
	\end{center}
\end{figure}

The results of the spectra for the Ta $4f$ core-level are shown in Fig. \ref{xpstao}(a). All spectra exhibit the typical spin-orbit doublet corresponding to the levels $4f_{7/2}$ and $4f_{5/2}$ located at 27.8 eV and 29.6 eV with peak splitting of 1.9 eV. The binding energies and peak splitting are characteristic of Ta$^{5+}$ in stoichiometric amorphous Ta$_2$O$_5$ films \cite{xmwu93a,kchen97a,amuto94a,rsimpson17a}. A closer look at the black spectrum for the as-deposited film (F298) reveals a satellite on the low energy region. This feature has been reported to be caused by screening of $5d$ electrons in $a$-TaO$_x$ ($x=1.86,2.00$), thus suggesting the presence of Ta suboxides such as TaO$_x$ \cite{ttsuchiya11a}. Since this film was not exposed to a heat treatment, one would expect, besides amorphous Ta$_2$O$_5$, the appearance of other Ta suboxides that were mainly generated during deposition as there still exists residual oxygen in the deposition chamber.

As the annealing temperature is increased, the shoulder vanishes indicating that the amount of suboxides is reduced.

\begin{figure}[t!]
	\begin{center}
		\includegraphics[width=8.4cm]{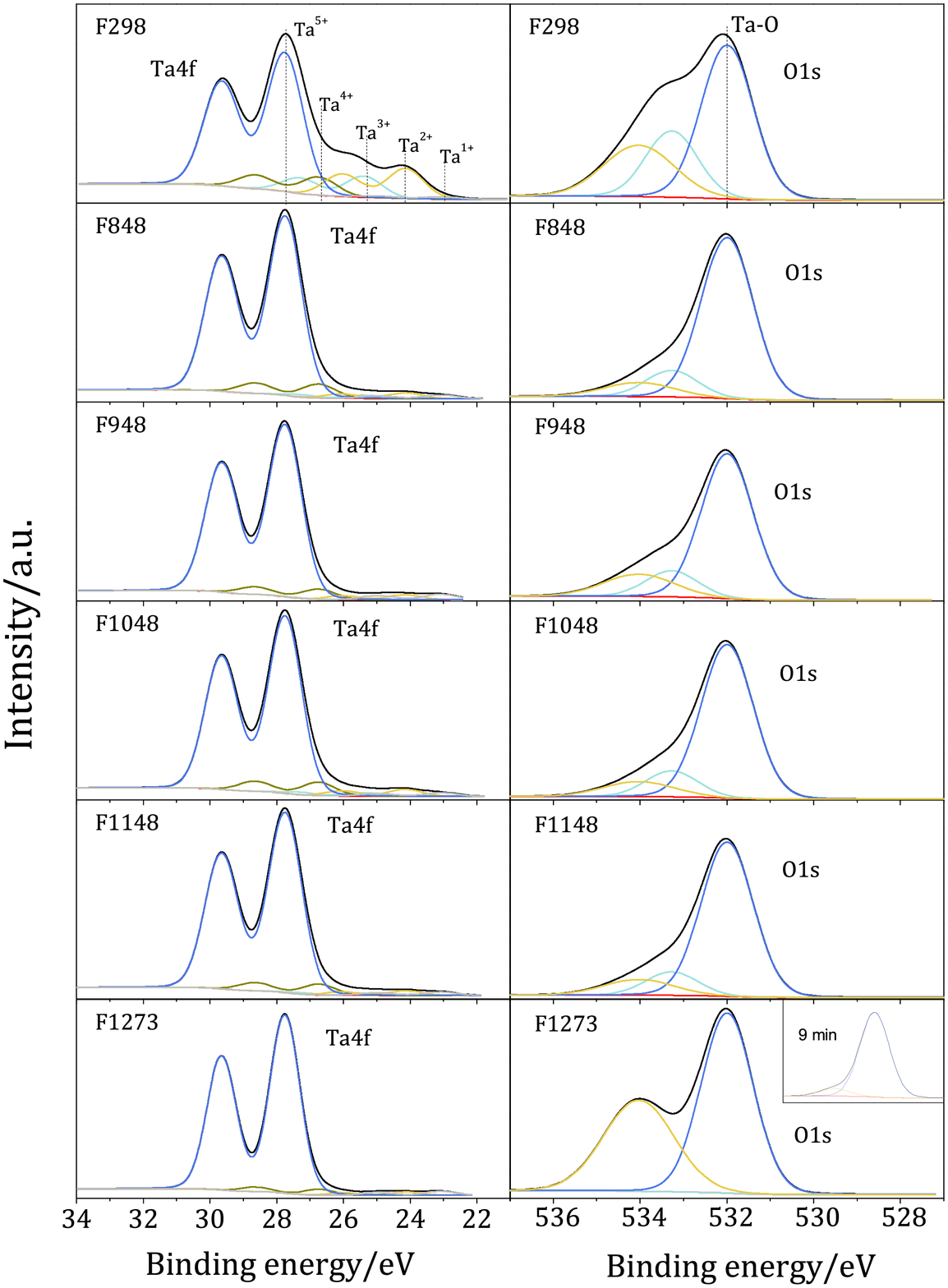} 
		\caption{Deconvolution of the Ta $4f$ (left column) and O $1s$ (right column) core-levels for the six films. The inset in the O $1s$ spectrum of F1273 shows the O $1s$ spectrum  after 9 min of sputtering. One can see that the satellite almost disappears and it is therefore caused by contamination.}
		\label{xpstaodeconv}
	\end{center}
\end{figure}

The O $1s$ core-level gives additional information on the chemical states of our samples; the corresponding spectra are shown in Fig. \ref{xpstao} (b). We first observe a peak at 532 eV associated to Ta-O bonds \cite{amuto94a,hszymanowski05a}. At $\sim$534 eV there is a satellite whose intensity reduces as $T_{ann}$ decreases. This feature is attributed to residual oxygen and surface contamination; mainly carbon compounds (see below). 
We also notice that for the sample F1273, the shoulder reappears. To verify the source of this shoulder we sputtered the sample for 9 min with an argon ion beam with a voltage of 3 kV, and an electric current of 10 $\mu$A ---the incidence angle between the sample and the ion gun was 90$^\circ$. Then the O $1s$ core-level was measured and the shoulder considerably diminished (as seen in the inset of the corresponding film in Fig. \ref{xpstaodeconv}). We thus believed that this feature is caused by contamination.

The deconvolution analysis performed on the Ta $4f$ spectra reveals that there are five contributions from five Ta oxidation states (see left column in Fig. \ref{xpstaodeconv}). The five doublets are located, with respect to Ta $4f_{7/2}$, at binding energies of (23.0, 24.1, 25.4, 26.7, 27.8) eV with spin-orbit splittings of 1.9 eV. These binding energies are attributed to Ta$^{1+}$, Ta$^{2+}$, Ta$^{3+}$, Ta$^{4+}$, and Ta$^{5+}$ states, respectively \cite{eatanassova98a,ivanov11a, ivanov11b}. These results strongly suggest the formation of metastable phases of Ta oxides such as TaO, Ta$_2$O, TaO$_2$, TaO$_{x}$, and Ta$_2$O$_3$. However, since the XRD show no traces of crystalline phases of Ta suboxides we believe that, due to the unstable nature of these phases, during annealing the surface atoms readsorbed O to form Ta$_2$O$_5$ and the other phases are trapped in a few topmost layers \cite{eatanassova03a}. On the other hand, the deconvolution of the O $1s$ spectra is shown on the right column of the same figure. There we can see that the oxygen spectrum is well fitted with three peaks. The low energy peak is attributed to Ta-O bonding, the peak at 533 eV is due to residual oxygen and the peak at 534 eV is due to carbon contamination, most probably carbon dioxide which is quite ubiquitous in all samples \cite{eatanassova98a,ttsuchiya11a,okerrec98a}. As discussed above the sample F1273 exhibits a pronounced satellite whose intensity was reduced after sputtering the sample for 9 min. The deconvolution confirms that this satellite is due to contamination.

\begin{figure}[t!]
	\begin{center}
		\includegraphics[width=8.4cm]{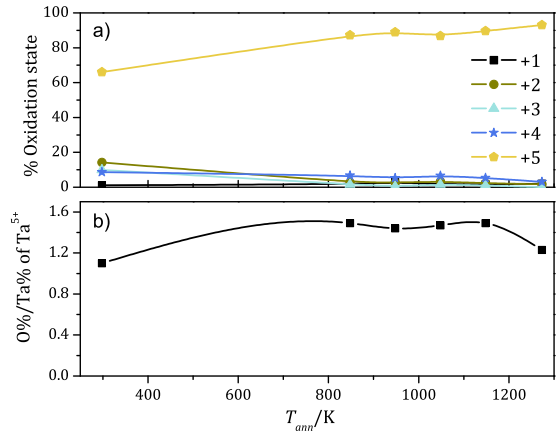} 
		\caption{Annealing temperature dependence of the oxidation state (a) and of the oxygen to tantalum ratio (b).}
		\label{percent}
	\end{center}
\end{figure}

In order to assess the effect of the annealing temperature on the chemical properties, we used the doublets of the Ta $4f$ spectra to compute the percentage of oxidation state as function of $T_{ann}$. Figure \ref{percent}(a) shows the results. We can observe a monotonic increase of the percentage of Ta in the state Ta$^{5+}$ as a function of annealing temperature. The increase for the state Ta$^{5+}$ goes from 67\% to 93\% whereas the rest of the states remain below 10\%. This tendency clearly demonstrates that the crystalline phases of Ta$_2$O$_5$ are favoured at high temperatures. We also computed the oxygen to tantalum ratio as a function of $T_{ann}$. The results are given in Fig. \ref{percent}(b). As expected, the film F298 exhibits an O\% to Ta\% ratio close to 1; nevertheless as the temperature is increased from 298 K to 1148 K, the ratio remains around 1.5 (non-stoichiometric) and decreases to 1.2 at 1273 K. This behaviour suggests that there is a range of temperature that favours the generation of tantalum pentoxide with the highest oxygen content. Accordingly, however, from 1148 K to 1273 K the oxygen seems to deplete. 

\section{Conclusions}
The atomic structure of Ta$_2$O$_5$ films was studied by X-ray diffraction. We found evidence for the presence of the hexagonal phase of Ta$_2$O$_5$ for samples annealed below 1048 K and the orthorhombic phase of Ta$_2$O$_5$ for samples annealed above 1048 K. In all these cases there are no traces of crystalline Ta suboxides. The films exposed to heat treatments below 848 K showed no diffraction patters and were found to be amorphous. 

XPS studies were realized in order to characterize the chemical properties of the films. According to the analysis carried out on the spectra for Ta $4f$ and O $1s$ core-levels, it can be concluded that Ta suboxides show up in small amounts as revealed by the appearance of satellites in both the Ta $4f$ and O $1s$ core-levels. The deconvolution of the XPS spectra strongly indicates the existence of several chemical states such as Ta$^{1+}$, Ta$^{2+}$, Ta$^{3+}$, Ta$^{4+}$, and Ta$^{5+}$. We conclude that as the annealing temperature is increased the presence of the state Ta$^{5+}$ increases, indicating that high annealing temperatures are desired to favor the generation of tantalum pentoxide. However, there seems to be a limit in the value of the annealing temperature if one wishes to avoid oxygen depletion. This is deduced by observing a reduction in the oxygen to tantalum ratio for the sample F1273.

\section*{Acknowledgements}
We are grateful to Wilian Cauich and Daniel Aguilar for their technical support during the XPS and XRD sessions. Dr. Israel Perez is indebted to Dr. Alberto Herrera for helpful discussions and technical support in the XPS analysis. We also thank the anonymous reviewer and one of the editors of this journal for their comments that greatly improved the quality of this work. The authors gratefully acknowledge the support from the National Council of Science and Technology (CONACYT) Mexico and the program C\'atedras CONACYT through project 3035.

\end{document}